\documentclass[reprint,aps,amsmath,amssymb]{revtex4}

\usepackage[]{graphicx}
\usepackage{subfigure}
\usepackage{times}
\usepackage{color}

\newcommand{\comment}[1]{}

\newcommand{\be}{\begin{equation}}
\newcommand{\ee}{\end{equation}}
\newcommand{\bea}{\begin{eqnarray}}
\newcommand{\eea}{\end{eqnarray}}

\begin{document}

\title{Transition-metal dichalcogenide heterostructure
 solar cells: A numerical study}

\author{A. Thilagam}
\email[]{thilaphys@gmail.com}

\affiliation{Information Technology, Engineering and Environment,\\ 
University of South Australia, Australia
 5095.}
\begin{abstract}
We evaluate the tunneling short-circuit current density  $J_{TU}$ in a  $p$-$i$-$n$ solar cell  in which the transition metal dichalcogenide heterostructure (MoS$_2$/WS$_2$ superlattice) is embedded in the intrinsic $i$ region. The effects of varying well and barrier widths, Fermi energy levels and number of quantum wells in the $i$ region  on $J_{TU}$
are examined. A similar analysis is performed for  the thermionic current $J_{TH}$ that arises due
to the escape and recapture of charge carriers between adjacent  potential wells in  the $i$-region. The interplay between $J_{TU}$ and $J_{TH}$ in the temperature range (300 K - 330 K) is examined.
The thermionic current is seen to exceed the tunneling current considerably at temperatures  beyond 310 K,
a desirable attribute in heterostructure solar cells.
 This work demonstrates the versatility of monolayer transition metal dichalcogenides
when utilized as fabrication materials for  van der Waals  heterostructure
solar cells.
\end{abstract}

\maketitle


\section{Introduction}
Low dimensional transition metal dichalcogenides (TMDCs) with the formula  MX$_2$
(M = Mo, W, Nb and X = S, Se)  are currently studied with 
great interest due to their potential applications in  
optoelectronics \cite{li2015heterostructures,li2016charge,perebeinos2015metal,ghatak2011nature}
 and photonics \cite{gan2013controlling,wu2014control,luo3few,
tsai2013few,eda2013two}. TMDCs possess
intra-layer metal planes  that are held between the  atomic planes of a  group VI element
by covalent bonding,  while  the inter-layers are weakly positioned by van der Waals forces.
Mechanical exfoliation using scotch tape techniques \cite{novoselov2005two} and more recent
 exfoliation techniques  \cite{fan2015fast,varrla2015large,chen2015nanoimprint}
allow convenient fabrication of quality monolayers  that display a  range of band-gaps within the visible spectrum. 
Monolayers of well known TMDCs (MoS$_2$,  MoSe$_2$, WS$_2$ and WSe$_2$)
 possess direct band-gaps \cite{makatom} unlike their
bulk counterparts that display  indirect band-gaps.
Due to the confinement of charge carriers, low dimensional
transition metal dichalcogenides 
possess unique photoluminescence properties that differ vastly from those in the bulk systems.
The widely studied molybdenum disulphide ((MoS$_2$) has an indirect bandgap 
of 1.29 eV  in bulk and is optically active in the visible
range with direct bandgap of 1.8 eV in the monolayer
form. The conduction band maximum and 
 valence band minimum (VBM) are  located at the $K$
symmetry points of the direct bandgap \cite{makatom,mak}.
The photoluminescence is enhanced even at room temperatures
with occurrence of notable spin-orbit splittings (0.15 $-$ 0.46 eV) 
due to a lack of  inversion symmetry 
at the  valence bands \cite{zhu2011giant}.

Transition metal dichalcogenide heterostructures  are designed  by stacking
at least two different  monolayers (MoS$_2$/WS$_2$, MoS$_2$/MoSe$_2$ 
or  WS$_2$/MoSe$_2$) in alternating series. These systems  possess  direct band gaps that range approximately
from 0.8 eV to 1.7 eV \cite{terrones2013novel,komsa2013electronic,
ionescu2015two,amin2015heterostructures,duesberg2014heterojunctions}.
The low dimensional TMDCs are based on the van der Waals interaction at atomically even interfaces where dangling bonds are noticeably absent. Consequently, 
lattice matching complications do not arise  giving rise to a wide choice in two-dimensional layer materials that can be used to fabricate  van der Waals  heterostructures.
Solar cells based on silicon on the other hand experience light-induced degradation
and  are limited by the indirect bandgap of
1.1 eV so that light absorption is restricted to the visible and near-infrared
range \cite{carlson1976amorphous,demichelis1985model,thilagam2001assessment}.
Monolayer TMDCs with thickness of less than 10 \AA \; 
 absorb up to 5 to 10\% incident sunlight,  one order of
magnitude higher than in GaAs and Si materials \cite{bernardi2013extraordinary}.
The optimized graphene/MoS2/$n$-Si solar cell system possess a
 conversion efficiency of 11.1\%  \cite{tsuboi2015enhanced} which is 
remarkably high for a  photovoltaic device derived from MoS$_2$ thin films.
A solar cell configuration based on  
vertically stacked Au/MoS$_2$/ITO PV devices yields
a high short-circuit photocurrent density of 20.9 mA/cm$^2$ and 
 power-conversion efficiencies of up to 2.8\% under
AM1.5G illumination \cite{wi2014enhancement}.
The  bandgap engineering possibilities of monolayer  TMDCs
 are thus desired for photovoltaic applications  \cite{howell2015investigation,wi2014enhancement,bernardi2013extraordinary}.

In the  WSe$_2$/WS$_2$ p-n heterojunction,  a type-II band alignment  \cite{huo2015tunable}
 results from the efficient separation of electrons and holes which is desirable for  enhanced photovoltaic effects. 
Similar physical separations of  electrons and holes  have also been observed  in the type-II  bilayers of 
alternative selenides and sulfides (WS$_2$/MoSe$_2$ and MoS$_2$/WSe$_2$)  \cite{terrones2013novel}.
Measurements from a combined photoluminescence spectroscopy and
optical pump-probe spectroscopy  show
direct evidence of charge transfer in photo-excited 
MoS$_2$/WS$_2$ heterostructures \cite{hong2014ultrafast}.
The time of  transfer of holes  in  the  MoS$_2$
layer to the WS$_2$ layer is estimated to be  less than
50 fs. First-principles density functional calculations 
confirm the occurrence of a type II MoS$_2$/WS$_2$ heterostructure \cite{kosmider2013electronic}
where the  electron and  hole
states at the optically active $K$ point are localized in different monolayers. 
The rapid rate of charge separation \cite{hong2014ultrafast}
is evidence that the high exciton binding energies
\cite{rama,makatom,chei12,komsa2012effects,thiljap,hill2015observation} 
does not present as an obstacle to  the transfer of   charge carriers in 
MoS$_2$/WS$_2$ heterostructures. The newly formed excitons appear to be ionized rapidly
into free electron-hole pairs due to a  built-in electric field
that is created when  carrier states are aligned differently in the  monolayers. The exact strength
of the internal field may be determined using
 computational techniques that probe the hybridized electronic states at the heterojunction interfaces and 
including details of interactions of photo-excited states 
with  lattice vibrational modes.

The ultrafast charge separation in heterostructures 
\cite{huo2015tunable,terrones2013novel,hong2014ultrafast,kosmider2013electronic}
appears to be a key factor that can be exploited to fabricate  solar cells with high  conversion efficiencies. 
This works examines whether solar cells that incorporate heterostructures
will give rise to enhanced  photovoltaic features.
The modification of the $i$ region to include a quantum well of a  narrow bandgap semiconductor 
 between two wide bandgap barrier semiconductors
 was first proposed by Barnham and Duggan \cite{barnham1990new}  so as to 
increase the photocurrent and to enhance the short-circuit current in solar cells.
A notable increase in the short circuit current and solar conversion efficiency
was seen in the case of GaAs/Al$_x$Ga$_{1-x}$As heterostructure solar cells \cite{barnham1991short}.

The  superlattice solar cell  is an extension of the single quantum well solar cell
in which   the  high bandgap and low bandgap   configuration of thin layers is repeated several times to enhance the tunneling  of  photo-generated carriers. 
The de-localization of charge carriers in distinct quantum wells depends
on the strength of coupling between the potential wells. In strongly coupled systems,
 the superlattice structures  form minibands which result in 
highly conductive systems with reduced recombination effects.
A study on the GaAs/Al$_x$Ga$_{1-x}$As superlattice solar cell \cite{aperathitis2001temperature},
showed a large output power under intense illumination compared to the
single quantum well solar cell. The efficiencies of solar cells based on 
the GaAs/Al$_x$Ga$_{1-x}$As quantum well structure 
are optimized at critical well widths \cite{thilagam1998optimizing}.
Similar studies have not been performed on transition metal dichalcogenides, hence
in this  work we examine a theoretical model 
in which the intrinsic $i$ region is made up of the MoS$_2$/WS$_2$ superlattice within the
$p$-$i$-$n$ solar cell system. Using numerical techniques we  investigate whether
the   photocurrent is enhanced  due to an increase in the number of  multi-layer TMDCs with 
direct band gaps in the  $i$ region. This study aims to
provide theoretical support  to the experimental realization of novel
solar cells based on  transition metal dichalcogenides  which enable
 enhanced photovoltaic attributes.
The MoS$_2$/WS$_2$ system is chosen for this study  as
experimental and theoretical data  \cite{hong2014ultrafast,kang2013band,kosmider2013electronic} of the type-II band structure are easily available for these materials. While a specific material combination is considered here, the approach used in this work can  be applied to other  monolayer  transition
 metal dichalcogenides \cite{kang2013band,ionescu2015two,amin2015heterostructures,duesberg2014heterojunctions} as well.

The paper is organized as follows. In
Sec. \ref{basic} we describe the basic operating principle
of the $p$-$i$-$n$ solar cell with the  MoS$_2$/WS$_2$
superlattice  inserted in the $i$ region. The general theory
of the tunneling currents in the solar cell is discussed, and
a relation for the tunneling current is used to examine the  
effects of varying well and barrier widths, Fermi energy levels and  temperature in the $i$ region.
Comparison of  the current estimates for a range of solar cell configurations
are made with existing experimental results as well.
In Sec.\ref{therm}, we examine the thermionic currents in the MoS$_2$/WS$_2$ superlattice $p$-$i$-$n$ solar cell and demonstrate the interplay between
tunneling and thermionic currents in the temperature range (300 K - 330 K).
We show that the thermionic current far
exceeds the tunneling current  at temperatures  beyond 310 K
which can be attributed to specific  processes which are highly sensitive
to the temperature. A brief discussion of  the mechanisms that degrade the solar cell efficiency
(trapping by surface defects, exciton recombination) is provided in Sec. \ref{fact}
and  conclusions are finally summarized in Sec.\ref{conc}.

\section{Numerical Analysis of the basic
$p$-$i$-$n$ solar cell \label{basic}}

\subsection{Basic model of the $p$-$i$-$n$ solar cell incorporating 
van der Vaals heterostructures}

\begin{figure}[htp]
  \begin{center}
    \subfigure{\label{figa}\includegraphics[width=7.85cm]{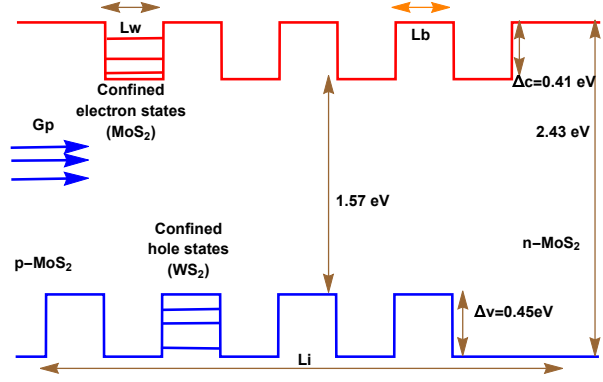}}\vspace{-1.1mm} \hspace{1.1mm} 
 \end{center}
     \caption{ Schematic structure of the $p$-$i$-$n$ solar cell incorporating the type-II MoS$_2$/WS$_2$ superlattice. The bandgap  differences (1.57 eV, 2.43 eV) are based 
 on  first principle calculations of the  conduction and valence band energy estimates
derived by Kang et. al. \cite{kang2013band}. The depths of the potential well are
quantified by $\Delta_c$ and  $\Delta_v$.  $G_p$ is the incoming photon flux, with  arrows
 on the left denoting the incoming photons. The width  of the intrinsic region is denoted by $L_i$.
}
 \label{cell}
\end{figure}

A simple layout of the  solar cell incorporating the vertically stacked 
 MoS$_2$/WS$_2$
superlattice  with a  number of repeated distances is shown in Fig.\ref{cell}. It  consists 
of a $p$-$i$-$n$ structure with alternating layers of   TDMCs (MoS$_2$ and WS$_2$) 
within  the $i$ region. Each layer of  MS$_2$ (M = Mo or W)
is made up of a plane of  Molybdenum or Tungsten atoms
sandwiched between two planes of sulfur atoms. The hybrid layers
are held by weak Van der Waals forces. 
The  spatial separation of  electrons and  holes gives rise to
the type-II band alignment of
 MoS$_2$/WS$_2$ heterostructure \cite{kosmider2013electronic}.
The valence band maximum lies in the WS$_2$ layer
while the  conduction band minimum lies in the MoS$_2$ layer.
Accordingly  the MoS$_2$/WS$_2$  bilayer heterojunction 
has an energy gap that is less than the band gaps of each of
 the two distinct monolayers. The monolayer MoS$_2$ presents as 
the barrier material for holes while monolayer WS$_2$ acts as
the barrier material for electrons. 


The  conduction and valence band energy estimates for the MoS$_2$/WS$_2$ heterostructure 
as indicated in Fig.\ref{cell}
are based on  first principle calculations \cite{kang2013band} of the many-body Schr\"odinger 
equation derived using density functional 
theory (DFT) and the Hartree-Fock (HF) approximations. Likewise
 the height of the electron and hole potential wells indicated in  Fig.\ref{cell}
are evaluated using estimates derived by Kang et. al. \cite{kang2013band}. These values
are based on 
the  Heyd-Scuseria-Ernzerhof  (HSE06) hybrid functional
which combines  Hartree-Fock and  density functional approaches.
We specifically utilize the HSE06 derived  energy results \cite{kang2013band} as these 
generally agree with experimental results \cite{hong2014ultrafast}
which show that the  monolayer MoS$_2$ and WS$_2$
possess bandgaps of 2.39 eV and 2.31 eV respectively. The valence band
maximum  in MoS$_2$ is 350 meV less than the valence band
maximum  in WS$_2$. 
Both $n$-type \cite{makatom,mak} and $p$-type \cite{laskar2014p} conductivities  occur in the monolayer MoS$_2$ 
depending on the material (e.g SiO$_2$ or Nb) on which it is deposited.
Here we assume  that the  $p$-region and $n$-region of the $p$-$i$-$n$ solar cell are derived using MoS$_2$, without too much focus on the details of the impurities in the $n$-type and $p$-type material
as the specific details of the end regions of the solar cell  do not affect the overall outcome of this study.

The heterostructure in the $i$ region is modeled as 
an alternating series of MoS$_2$/WS$_2$ with adjustable  widths considered for the well and barrier material in the 
range, $L_w$ =  $L_b$ = 10 to 14  \AA \;.
These widths  are justified on  the  basis of the Mo-W distance
of $d$ = 13 \AA \; over the vacuum region \cite{komsa2013electronic}  and
the monolayer MoS$_2$ height, $h$ =  9 \AA\; \cite{baugher2013intrinsic}.
The charge carriers occupy discrete
energy levels within the well regions where the energy gap is lower than that
of the barrier region due to quantum confinement
effects. The whole  system is subjected to an electric field
in the operating regime.

\subsection{Tunneling currents in the MoS$_2$/WS$_2$ heterostructure $p$-$i$-$n$ solar cell}

The presence of charge carriers outside the potential wells 
occurs with  non-zero probabilities that are based on quantum mechanical rules. The
tunneling process  is classically forbidden   for charge carriers which do not possess the extra
 energy required to escape the potential barriers.
Following the solar cell model of Aperathitis et. al. \cite{aperathitis2001temperature},
we consider that the confined charge carriers give rise to a 
 tunneling current ($J_{TU}$) in the growth direction of the device alongside a thermionic 
current ($J_{TH}$)  that occurs due to the escape of charge carriers from the quantum wells.
The  general current-density relationship for
current densities is given by $J = n q v$ where $q$ is the electron charge,
$n$ is the concentration of carriers with sufficient energy to overcome the potential
barriers and  $v(E)$  is the velocity of the carriers.
The potential wells shown in Fig\ref{cell}
act as electron or hole   reservoirs and influence
the net electron propagation from the p-region to the n-region
and hole propagation from the n-region to the p-region. 
The current flow is determined by 
the charge carrier  tunneling probability $P(E)$, 
density of states $g(E)$, Fermi-Dirac occupation probability $f(E)$
and  the photo-generation rate, $G_{ph} \; \alpha(\lambda) \; \exp(- \alpha(\lambda) x)$
where $\alpha(\lambda)$  is the absorption coefficient of the
well material.  In this work we consider that both the electron and 
hole contribute to the tunneling short-circuit current density due to the almost equivalent effective masses of electrons and holes in MoS$_2$,
and similar rates of transfer of holes and electrons between  the MoS$_2$ and  WS$_2$ layers.
 
The tunneling short-circuit current density  due to the charge carriers is expressed as 
\be 
\label{tuncur}
J_{TU}^i = q (\frac{A \tau}{L_i})  \int_0^{L_i} G_p \; \alpha(x) \; \exp[- \alpha(\lambda) x]  \; dx\; \int_{E_i}^{E_f} \;  g(E)\; v(E)
\; P(E) \; \frac{1}{1+ e^{ \frac{E-E_{F}}{k T}}}   \; dE,
\ee
where $i = e, h$ denotes current due to electron or holes,  $E$ is the confinement energy of the carrier in the potential well,
$\tau$ is the charge carrier transit time, $A$ is the area of cross-section of the solar cell, $L_i$ is the length of the intrinsic region.  The Fermi-Dirac distribution function described by $1/[1+\exp(\frac{E-E_{F}}{k_b \;T})]$
incorporates the temperature $T$ and  the quasi-Fermi levels $E_{F}$ = $E_{Fe}$ ($E_{Fh}$) which influences the propagation of electrons (holes).  The  Fermi level $E_{Fe}$ 
 lies within  the MoS$_2$ band gap and is pinned 
below the bottom of the conduction bands. The Fermi level $E_{Fe}$ for the  n-type MoS$_2$ is recorded
as -0.34 eV \cite{liu2011performance} which is marginally larger in magnitude than the 
Fermi level for   WS$_2$ at -0.33 eV. To simplify the model calculations, we assume a common Fermi
level for the MoS$_2$ and WS$_2$ layers.  The estimates of the  Fermi level $E_{Fh}$ for the  p-type MoS$_2$ \cite{dolui2013origin} is not available in the  literature. However the  Fermi level range (0.32 to 0.37 eV) is justified as one can employ
  mechanical bending processes  to control the Fermi-level shifts
in  two-dimensional materials \cite{yu2016bending} due to the
 localization of charges near the valence bands in MoS$_2$.
The influence of the Fermi level on the tunneling short-circuit current density
can thus be assessed by varying the Fermi levels $E_{Fe}$ and/or  $E_{Fh}$
to lie in the range 0.32 eV - 0.37 eV.

In Eq.\ref{tuncur}, we use  the carrier lifetime $t_c$ =  100 ps in 
MoS$_2$  \cite{wang2012ultrafast} to estimate the carrier transit time, $\tau = t_c/g$ = 1 ps at 
gain $g \approx$ 100.  The energy independent 2D density of states is given by $g(E) = \frac{g_s g_v m_i}{2 \pi \hbar^2}$ where
$g_s$=2 and $g_v$=2 are the respective spin and valley degeneracy factors, and
$\hbar$  and $m_i \; $ are the reduced Planck constant and effective mass of the electron ($m_e$)
 or hole ($m_h$).
To  simplify the numerical analysis,  we assume the effective mass to be an invariant of position in the heterostructure system. 
 The absorption coefficient for the MoS$_2$ depends on the incident
wavelength, and in this work we employ the empirical estimate of
$\alpha(\lambda)$ = 3 $\times$ 10$^5$ cm$^{-1}$  at the absorption wavelength
of 530 nm \cite{beal1979kramers}. The velocity  of the electrons can be
determined by employing $v(E) = \sqrt{2(\Delta_c-E)/m}$
where $\Delta_c$ is the conduction band discontinuity (shown in Fig \ref{cell}).  A similar relation
can be used to determine the velocity  of the holes using
 $\Delta_v$,  the valence band discontinuity.  For energies less
than the barrier height, $E < \Delta_c$, the 
 tunneling probability function  $P(E)$ associated with electrons (cf. Eq.\ref{tuncur})
is computed using
\be 
\label{prob}
P(E) = \left( 1+ \frac{\Delta_c^2 \sinh^2(k_b L_b)}{4 E (\Delta_c - E)} \right )^{-1}
\ee
where the potential barrier factor $k_b = (\sqrt{2 m_e (\Delta_c - E)}/\hbar)$.
The expression for $P(E)$ associated with holes can likewise be derived
using $\Delta_v$. The tunneling probability function  $P(E)$ in Eq.\ref{prob}
varies from the one employed by Aperathitis et. al. \cite{aperathitis2001temperature}
\be 
\label{prob2}
P_2(E) = \frac{16 E (\Delta_c-E)}{\Delta_c^2} \; \exp(-k_b' \sqrt{1-E/\Delta_c})
\ee
 where the potential barrier factor $k_b' = 2 L_b (\sqrt{2 m_e \Delta_c}/\hbar)$.
In this study, we evaluate the current density using the two different forms
of the  tunneling probability functions (Eq.\ref{prob}, \ref{prob2}) to make comparison
of the two functions using the  computed values of $J_{TU}$.

The  minimum (maximum) allowed energy in the quantum wells, $E_i$ ($E_f$) in Eq.\ref{tuncur}
 can be determined using the transcendental equations
\bea
\nonumber
\tan(\eta) &=& \sqrt{{\left(\frac{\sigma_0}{\eta} \right)}^2 -1} \; \quad {\rm (even \; parity)}, 
\quad 
-\cot(\eta) = \sqrt{{\left(\frac{\sigma_0}{\eta} \right)}^2 -1} \; \quad {\rm (odd \;  parity)} \\
\eta &=& \frac{L_w}{2} \sqrt{\frac{2 m_e}{\hbar^2} E}, \quad  \sigma_0 = \frac{L_w}{2} \sqrt{\frac{2 m_e}{\hbar^2} \Delta_c}
\label{evalues}
\eea
The energies $E_i$ and $E_f$  are quantified relative to the quasi-Fermi levels, $E_{Fe}$ or  $E_{Fh}$
which lie in the vicinity of the MoS$_2$ conduction or valence bands, respectively.
Substitution of  $P(E)$ in Eq.\ref{prob} into Eq.\ref{tuncur} will yield a 
simpler form of the tunneling current due to electron propagation 
\bea
\label{tcurrs}
J_{TU}^e = \frac{8 \sqrt{2} A \; q \; \tau \; g_s g_v \sqrt{m_e \;\Delta_c}}{\pi \hbar^2 L_i}\; G_p \; \left [1-\exp (- \alpha \;L_i)\right ] \; 
 \int_{E_i}^{E_f} \; (1-E/\Delta_c)^{1/2}\; \\ \nonumber
\times \left( 1+ \frac{\Delta_c^2 \sinh^2(k_b L_b)}{4 E (\Delta_c - E)} \right )^{-1} \; \frac{1}{1+ e^{\frac{E-E_{F}}{k T}}} \; dE.
\eea
A  relation similar to  Eq.\ref{tcurrs} is applicable for the
tunneling current due to hole propagation, $J_{TU}^h$.
The integral over the carrier energies in Eq.\ref{tcurrs}
is evaluated using the  minimum and maximum allowed energies $E_i$  and $E_f$
(cf. Eq.\ref{evalues}) based on given values of $A$, $G_p$ and  width  of the intrinsic region, $L_i$.  In  the 400-700 nm range of the visible region 
of the AM1.5G solar spectrum,
the  average incident photon flux is evaluated to be  $G_p$ = 1 $\times$ 10$^{17}$ cm$^{-2}$ s$^{-1}$.
The effect of the number of wells $n$ in the intrinsic region is examined by
setting $L_i = n\; (L_w + \;L_b) $. 
The junction area $A$ obviously determines the magnitude of the
current density $J_{TU}$.
In a study involving the GaAs/Al$_x$Ga$_{1-x}$As superlattice 
solar cell system  \cite{aperathitis2001temperature},
a small junction area $A$ = 4 $\times$ 10$^{-3}$ cm$^2$  gave rise to a tunneling 
short-circuit current density $J_{TU}$  in the range  3.5 to 43 mA/cm$^2$. This result
also depended on  device parameters ($L_w$, $L_b$) in the $i$-region and  temperature. 
In a recent work by  Tsai et. al. \cite{tsai2014monolayer},
a Al-contacted monolayer MoS2/$p$-Si solar cell of 
area  1 cm $\times$ 1 cm was fabricated  with doping concentration between 3.2 $\times$ 10$^{16}$ and 2.5 $\times$ 10$^{17}$ cm$^3$. This system
displayed the open circuit voltage $V_{oc}$ of 0.38 to 0.41 V and a short-circuit 
current $J_{sc}$ from 21.66 to 22.36 mA/cm$^2$ with an  efficiency of 5.23\%.
One needs to exercise  caution when comparing results as  solar cells of different
 configurations and dimensions are used in various studies.
In this work  a small value of $A$ = 1 $\times$ 10$^{-4}$ cm$^2$ is chosen as
we expect a short-circuit current that is comparable to experimental observations \cite{tsai2014monolayer}
due to    incorporation of the MoS$_2$/WS$_2$ superlattice in the solar cell system.

\begin{table*}
    \caption{\label{tab:gw} 
Parameters used to model  the $i$-region of the $p$-$i$-$n$ solar cell. The total current density 
$J_{TU} = J_{TU}^e + J_{TU}^h$ is evaluated using  Eq.\ref{tcurrs}, 
at temperature  $T$ = 300 K and is based on the 
tunneling probability function given in Eq.\ref{prob}.  $J_{TU}$ is given in units of mA/cm$^2$. 
The electron and hole energies ($E_i^e$, $E_i^h$) are calculated relative to the
bottom of the quantum well. The total width of intrinsic region is fixed at $L_i$ $\approx$ 0.1 $\mu$ m. The magnitude of  the  Fermi energy levels $E_{Fe}$ and $E_{Fh}$ are taken  to be  0.34 eV.
The effective mass of the electron $m_e$ = 0.51 $m_o$ (MoS$_2$) and 
hole $m_h$ = 0.42 $m_o$ (WS$_2$)  given in terms of the free-electron mass $m_o$
are retrieved from Ref.\cite{jin2014intrinsic}. The junction area, 
$A$ = 1 $\times$ 10$^{-4}$ cm$^2$ while the carrier transit time, $\tau$ = 1 ps. 
Based on the AM1.5G solar spectrum, the  average incident photon flux is evaluated to be  $G_p$ = 1 $\times$ 10$^{17}$ cm$^{-2}$ s$^{-1}$. The current density estimates included within the brackets below are evaluated using the tunneling probability function given in Eq.\ref{prob2}.
\newline }
\begin{tabular}{|c|c|c|c|c|c|c|c|c|c|c}
        \hline
        \hline
Well period & \; $L_w$\;(\AA) \; & \; $L_b$ \;(\AA) \; & \; $E_i^e$ (eV) \; & \; $E_i^h$ (eV) \; & \; $J_{TU}^e$  (mA/cm$^2$) \;& \; $J_{TU}^h$  (mA/cm$^2$) \; & \; $J_{TU}$  (mA/cm$^2$) \;\\         \hline
56 & 9  &  9 &  0.21 & 0.25& 12.42 \;(13.98) & 4.08 \; (4.68)& 16.50 \; (18.66) \\
50 & 10  &  10 &  0.20 & 0.22 & 18.49 \;(20.02) & 6.77 \; (7.46) & 25.26 \; (27.48) \\
 45 & 11 &  11 &  0.18 & 0.21 & 24.36 \; (25.66) & 9.92 \; (10.59)& 34.28 \; (36.25) \\
 42 & 12 & 12 & 0.16   & 0.19 & 28.30 \; (29.21) & 12.71 \; (13.26) & 41.01 \; (42.47) \\
 38 & 13 &  13 & 0.15 &  0.17 & 31.03 \; (31.68) & 15.29 \; (15.72) & 46.32 \; (47.40) \\
 36 & 14 &  14 & 0.14 &  0.16 & 30.64 \; (31.04) & 16.48  \; (16.78) & 47.12 \; (47.82)  \\
30 & 16.5 &  16.5 & 0.11 &  0.13 & 24.63 \; (24.74) & 16.00 \; (16.09) & 40.63 \; (40.83) \\
25 & 20 &  20 & 0.09 &  0.10 & 11.84 \; (11.85) & 9.52  \; (9.53) & 21.32 \; (21.38)  \\
        \hline
        \hline
    \end{tabular}
\end{table*}

\begin{figure}[htp]
  \begin{center}
\subfigure{\label{figa}\includegraphics[width=7.85cm]{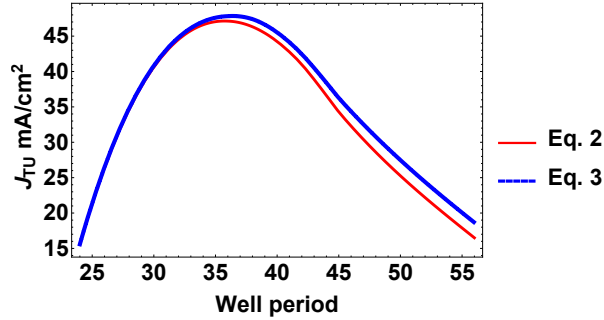}}\vspace{-1.1mm} \hspace{1.1mm} 
 \end{center}
     \caption{ The total current density 
$J_{TU} = J_{TU}^e + J_{TU}^h$ (mA/cm$^2$) as function of the well period 
at temperature 300 K and Fermi energy levels $E_{Fe}$ = $E_{Fh}$ = 0.34 eV.
All other parameters, including the variables $L_w$ and  $L_b$,   
are the same as those used to obtain the results in  Table-I. The estimates are computed
using Eq.\ref{tcurrs},  the tunneling probability function  $P(E)$ (Eq.\ref{prob}, red solid) and $P_2(E)$ (Eq.\ref{prob2}, blue solid).}
 \label{well}
\end{figure}

The results in Table-I obtained using the specified parameters for the $i$-region 
 illustrate the importance of incorporating quantum wells  within the $p$-$i$-$n$ solar cell.
At a constant  width $L_i$ of the intrinsic region, 
the total current density $J_{TU}$ increases with well period 
 before a decrease occurs beyond an optimum well period (about 36)  with $L_w$ = $L_b$ = 14 \;\AA.
There are competing factors which account for the peak in current density observed at 
the optimum well period shown in Fig.\ref{well} for the parameters specified in Table-I.
An  increase in the number of potential wells in $i$-region enhances the current density.
On the other hand, a larger number of wells correspond to 
 decreased well width  for which the  minimum energy levels
of the confined charge carriers are increased. This in turn lowers the  
tunneling probability through the $i$-region  of the $p$-$i$-$n$ solar cell
 restricting further rise in   $J_{TU}$.  We note that the total current density $J_{TU}$ 
evaluated using Eq.\ref{tcurrs} in Table-I is  of the same order in magnitude as the experimental short-circuit current  density of 33.4 mA/cm$^2$ \cite{tsuboi2015enhanced}
derived from the trilayer-graphene/MoS2/n-Si solar cell with a 9 nm MoS$_2$ film. This optimized solar cell
configuration resulted in  the photovoltaic efficiency of 11.1\% \cite{tsuboi2015enhanced}. Although direct
comparison of results is challenging  due to differences in solar cell configurations,
our results nevertheless demonstrate the potential for the MoS$_2$/WS$_2$ superlattice $p$-$i$-$n$ solar cell to produce equivalent or larger photovoltaic efficiencies under favorable operating conditions.


\begin{figure}[htp]
 \begin{center}
\subfigure{\label{figa}\includegraphics[width=7.25cm]{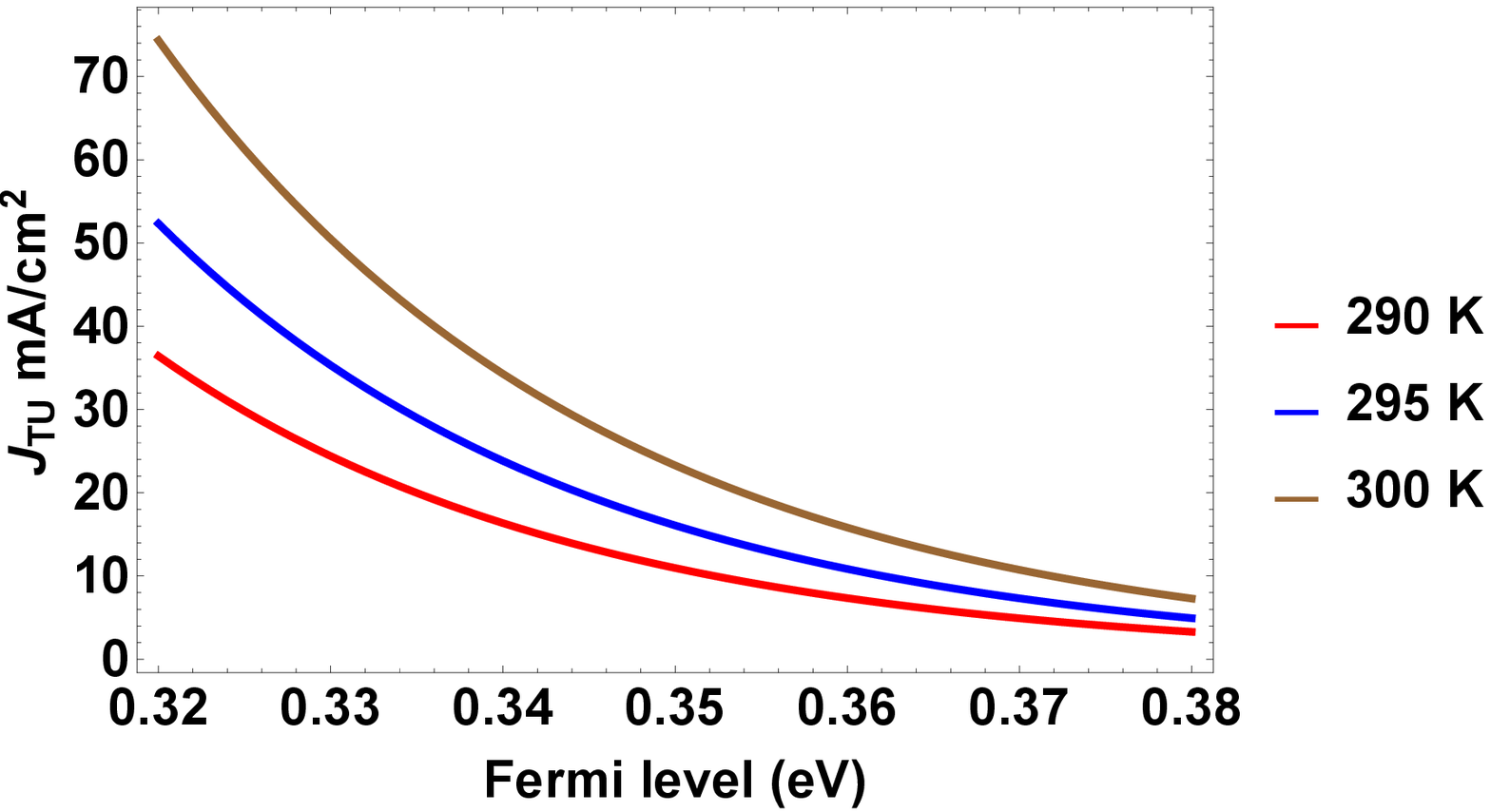}}\vspace{-1.1mm} \hspace{1.1mm} 
\hspace{10mm}
\subfigure{\label{figb}\includegraphics[width=5.95cm]{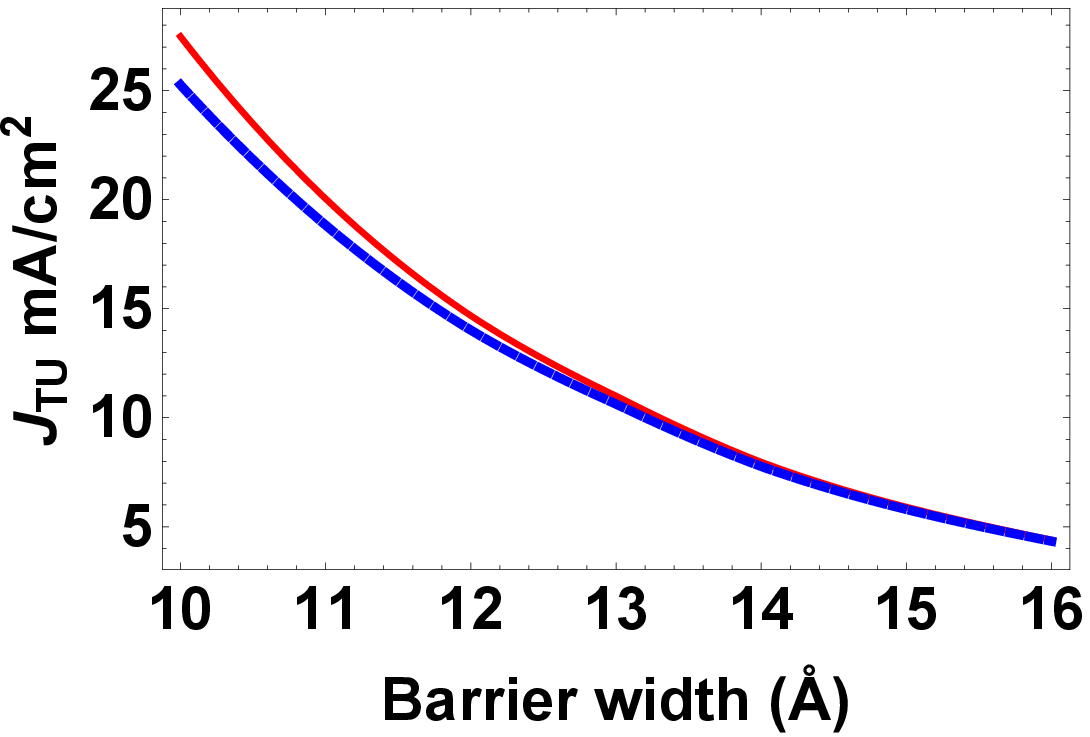}}\vspace{-1.1mm} \hspace{1.1mm} 
\end{center}
     \caption{ (a) The total current density 
$J_{TU} = J_{TU}^e + J_{TU}^h$ (mA/cm$^2$) as function of the Fermi  levels $E_{Fe}$ = $E_{Fh}$ 
at temperatures (290 K, 295 K, 300 K) computed using Eq.\ref{tcurrs} and an analogous
relation for holes.  The well period = 45 and $L_w$ = 11 \AA,  $L_b$ = 11\;\AA. All other parameters  (except for the Fermi energy levels) are  given in  Table-I.\\
(b)  The total current density $J_{TU}$  computed using the
tunneling probability function  $P(E)$ (Eq.\ref{prob}, lower blue curve)
and $P_2(E)$ (Eq.\ref{prob2}, upper red curve).
The well width  is fixed at $L_w$ = 10 \;\AA \; and the barrier width $L_b$ is varied
such  that the total width of intrinsic region remains constant at $L_i$ = $n\; (L_w + L_b)$ $\approx$ 0.1 $\mu$ m
All  other parameters are the same as those used to generate the results of Table-I.}
 \label{fermi}
\end{figure}

The effect of the Fermi energy levels $E_{Fe}$ and $E_{Fh}$ on the total
current density $J_{TU} = J_{TU}^e + J_{TU}^h$  evaluated using the
tunneling probability function  $P(E)$ (Eq.\ref{prob})
is shown in Fig. \ref{fermi}a.
The well period = 45, $L_w$ = 11 \;\AA \;,  $L_b$ = 11\;\AA \;
  and  other parameters are the same as those used to generate the results of Table-I.
A significant increase in $J_{TU}$ occurs when the Fermi  levels 
are positioned closer to the conduction or valence bands. At the higher temperature of 300 K,
$J_{TU}$ is enhanced by a factor of 7 when the Fermi levels
are shifted closer to the conduction/valence bands by 0.32 eV from 0.38 eV .  
These results may  be linked to the lower ``activation energies"  that is needed for the confined charge carriers to tunnel through the potential barriers.


The wave-function of the charge carrier with larger mass experiences  
faster decay in the  barrier material compared to the carrier with
a lighter mass. The energy estimates in Table-I show that  the minimum energy levels 
of the electron with higher mass (0.51 $m_o$)  is lower than the energy levels of the hole (0.42 $m_o$).  The minimal energy level appear to be a dominating factor, as
it influences the tunneling probability through the $i$-region, accordingly
the current density $J_{TU}$ due to electrons is higher than that due to holes
for the parameters listed in Table-I. 
The transmission probability of a trapped charge carrier also decreases with the
thickness of the barrier  as shown in  Fig. \ref{fermi}b
where  the total current density $J_{TU}$ is computed using the
tunneling probability function  $P(E)$ (Eq.\ref{prob})
and $P_2(E)$ (Eq.\ref{prob2}) for  comparison purposes.
The well width  is fixed at $L_w$ = 10 \;\AA \; and the barrier width $L_b$ is varied
such that the total width of intrinsic region remains constant at $L_i$ = $n\; (L_w + L_b) $$\approx$ 0.1 $\mu$ m.
The other parameters used to obtain the results in Fig. \ref{fermi}b
are given in Table-I.
A slightly elevated $J_{TU}$ is obtained  with  the
tunneling probability function   $P_2(E)$ (Eq.\ref{prob2}) at barrier widths $\approx$ 10 \AA\;,
 otherwise the two probability distributions yield almost similar estimates of 
the current density. The results in Fig. \ref{fermi} show that $J_{TU}$ is  enhanced in a 
solar cell with  decreased barrier widths and  at elevated temperatures, the latter two factors contributing
to the  increased energies of charge carriers which enhances the tunneling effects.

\section{\label{therm} Thermionic currents in the MoS$_2$/WS$_2$ superlattice $p$-$i$-$n$ solar cell }

A  thermionic current \cite{wu1979carrier,anwar1999effects,allain2015electrical,
shichijo1980real,mottet1988thermionic,grinberg1986thermionic,
grupen1991simulation} is induced when charge carriers escape from
one quantum well and are recaptured by the adjacent well without undergoing 
recombination processes in  the $i$-region.
 The  thermionic emission  is an important mechanism that induces current
across the interface layers of the MoS$_2$/WS$_2$ system.
It  is evaluated based on the positions  of the conduction and valence bands and  changes in effective masses
of  the charge carriers that transverse  quantum mechanically through
the series of potential barriers. In order to obtain estimates of  the thermionic current for 
the MoS$_2$/WS$_2$ superlattice $p$-$i$-$n$ solar cell,
we consider the thermionic theory of Wu et. al. \cite{wu1979carrier}
for bulk systems  as well as its adaptation to two-dimensional systems
 \cite{anwar1999effects,allain2015electrical,wu1979carrier}.
We employ the  thermionic current of the   form
\be
\label{thermcu}
J_{TH} = \frac{\sigma}{L_q} A_r \; T^\alpha \sqrt{\frac{m_{e1} m_{e2}}{m_{e1}+m_{e2}}} \exp \left(- \frac{E_b-E_{Fe}}{k_B T} \right) \\
\ee
where $A_r$ = $ \frac{\sqrt{2 \pi} e k_b^\alpha}{\hbar^2}$  is  the 
Richardson constant \cite{crowell1965richardson} at $\alpha$ = 3/2,
$L_q$ = $L_b$ + $L_w$ is the quantization length  and $\sigma$
is a  material dependent correction factor which is $\approx$ 0.5 for a wide range of material
systems. The factor $\alpha$ = 2 for bulk materials and is $\frac{3}{2}$ for
two-dimensional systems \cite{anwar1999effects,allain2015electrical} and
$E_{Fe}$ is the Fermi energy level that lies in the vicinity of the conduction bands, 
and  $m_{e1}$ and $m_{e2}$ are the respective electron masses
in the  MoS$_2$ and WS$_2$ monolayer regions.  We set effective potential barrier energy 
$E_b$ = $\Delta_c-E_1$
where  $\Delta_c$ is the conduction band discontinuity (see Fig \ref{cell})
and $E_1$ is the   minimum  energy in the quantum wells. We employ the values
of $E_i$ evaluated for the specific cell configuration as detailed in Table-I. The  minimum  energy
$E_1$ was not included in  the computation of the thermionic current
in earlier works \cite{wu1979carrier,hojfeldt2002modeling,anwar1999effects,allain2015electrical}.
The inclusion of this minimum energy  adds greater accuracy to the evaluation of 
 the effective barrier via $E_b$ = $\Delta_c-E_1$.
The thermionic current due to holes appears in  the same form as Eq.\ref{thermcu}, with
substitution of the valence band discontinuity $\Delta_v$ and 
$m_{h1}$ and $m_{h2}$,  the respective hole masses
in the WS$_2$ and MoS$_2$ monolayers.

A standard form for the thermionic current (Eq.\ref{thermcu}) is lacking
in the literature, partly due to 
 inconsistencies in the definition employed for the   Richardson constant, 
$A_r$ \cite{anwar1999effects,allain2015electrical}.
Moreover the correction factor $\sigma$ in Eq.\ref{thermcu}
has not been considered in earlier works. A reflection factor which
takes into account backscattering processes and interactions with
secondary electrons \cite{modinos1984secondary} is expected to contribute
to a more realistic modeling of the thermionic current.
In order to obtain quantitative estimates of the thermionic current $J_{TH}$ 
using  Eq.\ref{thermcu}, we employ  $\sigma$ = 0.5 and  consider the masses of the charge carriers in the MoS$_2$ and WS$_2$ monolayers as  
$m_{e1}$ = 0.51, $m_{e2}$ = 0.31, $m_{h1}$ = 0.42, $m_{h2}$ = 0.58 \cite{jin2014intrinsic}.
In Fig. \ref{therm}a we plot the thermionic current $J_{TH}$ as a function of the
temperature based on  the minimum  energies $E_1$ for  electrons and holes 
(Table-I) and the common Fermi energy levels $E_{Fe}$ = $E_{Fh}$ = 0.34 eV
The results indicate that the thermionic current is  dominant in systems
with small potential well widths, rising rapidly with temperature for all system configurations. 

As  different formalisms are used to estimate the tunneling 
and thermionic currents in this study, we use the ratios
${J^r}_{TH}$ = $\frac{J_{TH}(T)}{J_{TH}(280 K)}$
${J^r}_{TU}$ = $\frac{J_{TU}(T)}{J_{TU}(280 K)}$
 to compare the rise of the different current components with temperature.
Fig. \ref{therm}b demonstrates the interplay between
tunneling and thermionic currents in the temperature range (300 K - 330 K).
While the difference between the currents is not significant in the
temperature range (270 K - 290 K), the thermionic current far
exceeds the tunneling current  at temperatures  beyond 310 K.
This can be attributed to  the higher probabilities of  charge carriers
with increased energies that escape from 
the potential well, which is also quantified by the  exponential function in Eq.\ref{thermcu}.
To this end, the thermionic current plays an important role 
in enhancing the performances of  heterostructure solar cells at elevated temperatures.

The combined results in Figs. \ref{well} and \ref{therm} 
show that the output currents in a heterostructure solar cell
can be  controlled to a desirable level
 by adjusting  the  dimensions of the  potential well and  barrier
widths of the superlattice in the $i$ region. The Fermi energy levels
also have a strong influence on the total current density ($J_{TU}$+$J_{TH}$).
The quantitative details of the influence of the depth of potential wells on the current 
densities have not been examined here. In general,
 deep wells (large $\Delta_c$, $\Delta_v$ in Fig. \ref{cell})  tend to hold
 a higher density of carriers  than  shallow wells,  which 
 contributes to increased current densities. Further investigation is 
needed to examine if deep wells impede the thermionic process as it is likely that the recapturing of emitted charge carriers becomes less efficient when the well depths  $\Delta_c$, $\Delta_c$ are made larger.


\begin{figure}[htp]
  \begin{center}
\subfigure{\label{figa}\includegraphics[width=8.2cm]{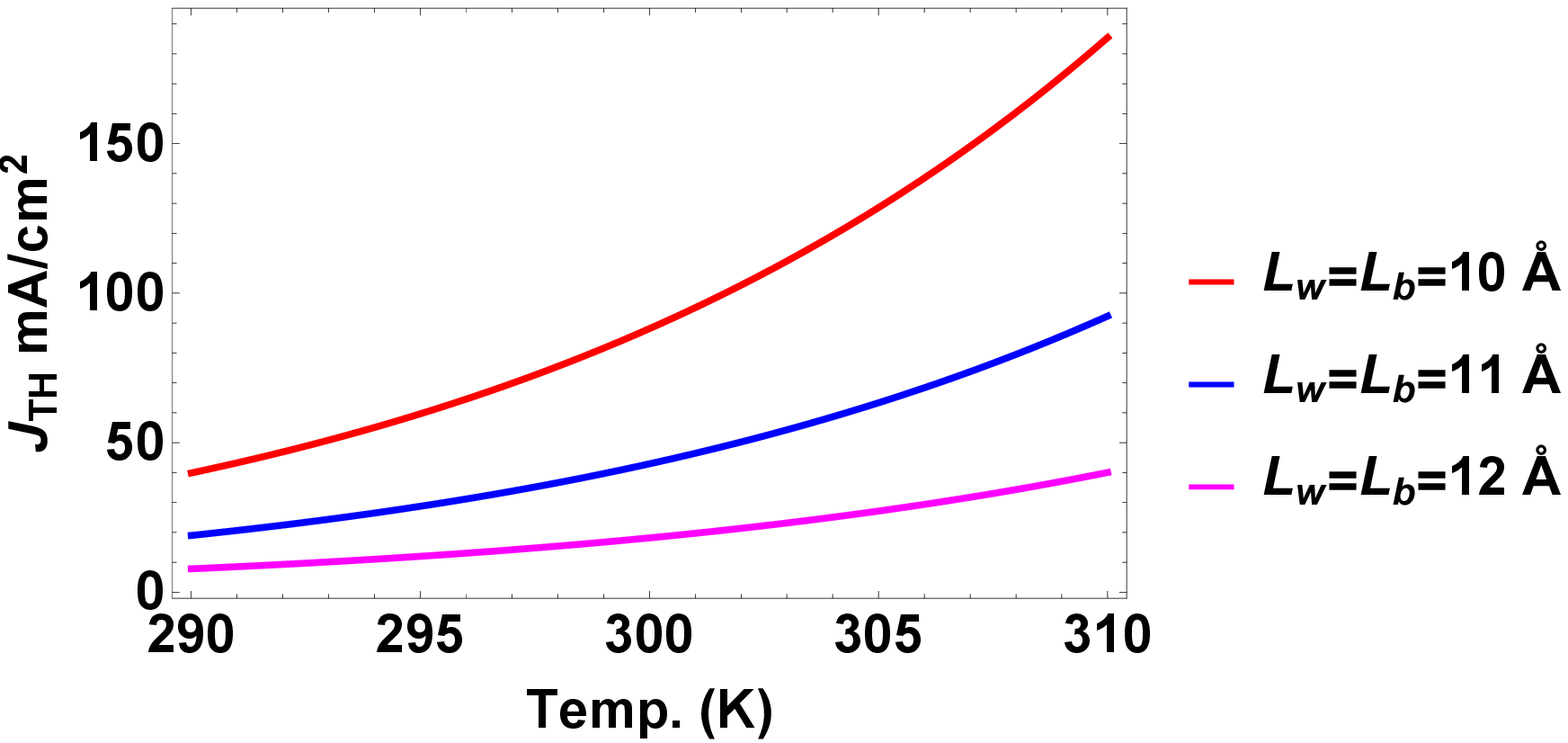}}\vspace{-1.1mm} \hspace{1.1mm} 
\hspace{10mm}
\subfigure{\label{figa}\includegraphics[width=7.4cm]{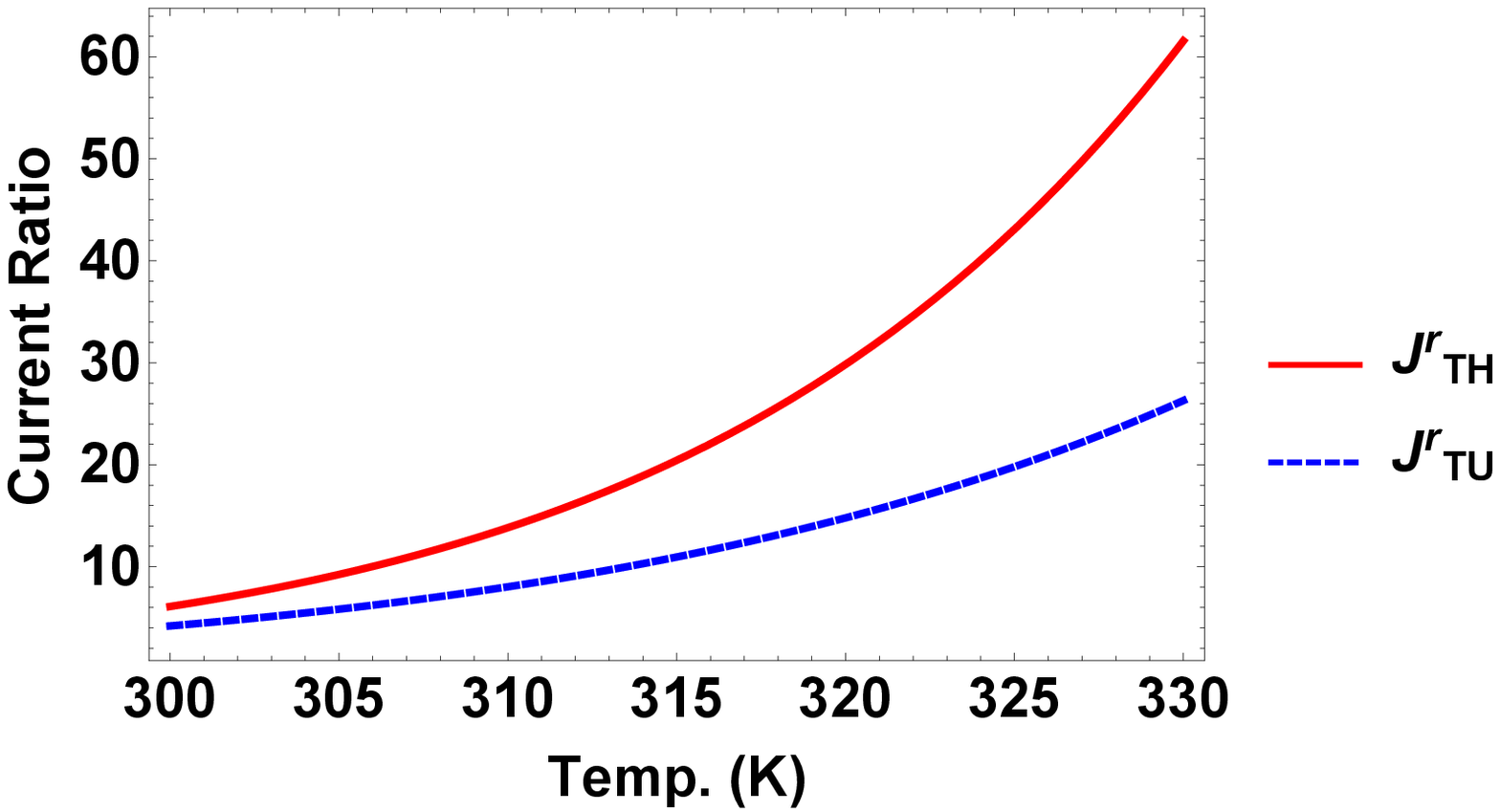}}\vspace{-1.1mm} \hspace{1.1mm} 
 \end{center}
     \caption{(a) The thermionic current density 
$J_{TH}$ (Eq.\ref{therm}) as function of the temperatures at three different well and barrier widths.  We set  $\sigma$ = 0.5 with masses of the charge carriers in the MoS$_2$ and WS$_2$ monolayers taken as $m_{e1}$ = 0.51, $m_{e2}$ = 0.31, $m_{h1}$ = 0.42, $m_{h2}$ = 0.58 \cite{jin2014intrinsic}. The minimum  energies $E_1$ for  electrons and holes are given in
Table-I and the common Fermi energy levels $E_{Fe}$ = $E_{Fh}$ = 0.34 eV 
 \\
(b) Comparison of the rise of the tunneling and thermionic currents with temperature based on the
ratios, ${J^r}_{TH}$ = $\frac{J_{TH}(T)}{J_{TH}(T_o)}$
${J^r}_{TU}$ = $\frac{J_{TU}(T)}{J_{TU}(T_o)}$ where $T_o$ = 280 K.
The widths $L_b$ = $L_w$ = 14 \AA \;, and $E_i^e$ = 0.14 eV,  $E_i^h$ = 0.16 eV are used
in Eqs. \ref{tcurrs} and \ref{thermcu} to compute the current densities.
}
 \label{therm}
\end{figure}

\subsection{Factors that  decrease the conversion efficiency in
 the $p$-$i$-$n$ heterostructure solar cell \label{fact}}

The  conversion efficiency quantifies the fraction  of solar energy that is
transformed via photovoltaic action into electricity. It is dependent
on several parameters  \cite{landsberg1993band,nozik2002quantum,wolf1998solar}
which have  not been actively examined in this work.
The appearance of a peak in current density $J_{TU}$
(Fig. \ref{well}) however implies that the conversion efficiency is not
  a  monotonic function of the number of potential wells in a  heterostructure solar
cell. The fast ionization of excitons  enhances $J_{TU}$ and contributes to increased
efficiencies  in  heterostructure solar cells. However excitons  undergo fast relaxation due to interactions
with phonons \cite{thilrelaxjap} and may recombine radiatively 
instead of undergoing ionization into free carriers.
The defect assisted scattering and  trapping of excitons 
by surface  states \cite{shi2013exciton} may also act to 
 decrease the efficiencies of layered transition  metal dichalcogenides 
solar cells. The  fast capture  of excitons   by mid-gap defects through
 Auger processes  \cite{wang2015fast}  and exciton-exciton annihilation \cite{sun2014observation,konabe2014effect} are alternative processes that may result in  decreased efficiencies of solar cells.

The formation of the negatively charged trion which is a   three-body bound state
of two electrons and one hole \cite{thiltrion} is expected to interfere with photoconductive properties
of the solar cell.  The charged exciton  in layered transition  metal dichalcogenides  are highly stable even at room temperatures  \cite{lee2016identifying,berkelbach2013theory,thiljap}.  
The high effective mass of the trion give rise to lower mobilities with decreased
 probabilities of tunneling through the potential barriers in the $i$-region of the
solar cell. To this end, the formation of charged exciton complexes is expected
to reduce solar cell efficiencies. A detailed examination of the expected
decrease in efficiency due to creation of trion quasi-particles in 
solar cells fabricated using transition-metal dichalcogenides is planned for future investigations.

\section{Conclusion \label{conc}}

In summary, we have investigated the importance of embedding the MoS$_2$/WS$_2$ superlattice
in the intrinsic $i$ region of a  $p$-$i$-$n$ solar cell structure.
A type-II band alignment of MoS$_2$ and WS$_2$ acts as the
driving force for  the effective charge separation and enabling
enhanced performances for  photovoltaic applications.
Our results show that the  tunneling short-circuit current density $J_{TU}$ increases with
the incorporation of quantum wells in the intrinsic $i$ region.
The tunneling  current density is dependent on  the
potential well width, barrier width,  Fermi energy levels and temperature. A decrease in the
 short-circuit current density  occurs when the
optimum number of quantum wells is exceeded in the $i$ region when
the overall width of the intrinsic region is held constant. 
The positioning of Fermi energy levels closer to the conduction or valence bands is seen to give rise to
a marked increase in $J_{TU}$. 

The effects of varying well and barrier widths and  Fermi energy levels 
on the thermionic current $J_{TH}$ show a trend similar to that displayed by the
tunneling current.  The interplay between $J_{TU}$ and $J_{TH}$ in the temperature range (300 K - 330 K)
is analysed. The thermionic current is seen to  far exceed the tunneling current  at temperatures  beyond 310 K. We emphasize the prominent role played by the  thermionic current 
in enhancing the performances of  heterostructure solar cells at elevated temperatures.
Our  results predict that the heterostructure solar cell is better suited to yield
higher current densities compared to homogeneous p-i-n solar cells made of 
transition-metal dichalcogenides.   It would be worthwhile to see if
experimental studies can produce results which are consistent with our theoretical findings.
In future investigations, the various loss mechanisms (exciton recombination, exciton-exciton annihilation) and polaronic effects \cite{matos2002fractional,thilagam2004excitonic,thilagam2015excitonic} may need to be  incorporated for a more realistic modeling of the heterostructure  solar cells.  

Lastly, the charge separation resulting in  electron confinement to the MoS$_2$ layer and hole confinement to the WS$_2$ layer  typifying the  type-II band alignment 
at the MoS$_2$/WS$_2$ interface presents a unique functionality that can be  exploited in
heterostructure  devices.  The results of this study while focussed 
on the MoS$_2$/WS$_2$ system  points to  new directions for 
novel solar cell designs based on alternative  combinations
of  MoSe$_2$, MoTe$_2$  and WSe$_2$.  
 Innovative   solar cell configurations that incorporate vertical or lateral heterostructures \cite{terrones2013novel,gong2014vertical} are  expected to replace
crystalline silicon solar cells which currently are limited by their maximum achievable
solar conversion efficiencies.

\end{document}